\documentclass[11pt,aps,amsmath,amssymb,nofootinbib,notitlepage,longbibliography]{revtex4-1}
\usepackage{amsmath,amssymb}
\usepackage{slashed}
\usepackage{graphicx}
\usepackage{bm}
\usepackage[top=1.0in,bottom=1.0in,left=1.0in,right=1.0in]{geometry}
\usepackage[colorlinks,linkcolor=blue,citecolor=blue]{hyperref}

\newcommand{\be}{\begin{equation}}
\newcommand{\ee}{\end{equation}}
\newcommand{\bea}{\begin{eqnarray}}
\newcommand{\eea}{\end{eqnarray}}
\newcommand{\ba}{\begin{eqnarray}}
\newcommand{\ea}{\end{eqnarray}}

\begin{document}

\title{Spatial entanglement in two dimensional QCD:\\
Renyi and Ryu-Takayanagi entropies}


\author{Yizhuang Liu}
\email{yizhuang.liu@uj.edu.pl}
\affiliation{Institute of Theoretical Physics, Jagiellonian University, 30-348 Kraków, Poland}

\author{Maciej A. Nowak}
\email{maciej.a.nowak@uj.edu.pl}
\affiliation{Institute
of Theoretical Physics and Mark Kac Center for Complex Systems Research,
Jagiellonian University, 30-348 Kraków, Poland}

\author{Ismail Zahed}
\email{ismail.zahed@stonybrook.edu}
\affiliation{Center for Nuclear Theory, Department of Physics and Astronomy, Stony Brook University, Stony Brook, New York 11794--3800, USA}



\begin{abstract}
We  derive a general formula for the replica partition function in the vacuum state,
for a large class of interacting theories with fermions, with or without gauge fields, using the equal-time  formulation
 on the light front. The result is used  to analyze the spatial entanglement of interacting Dirac fermions in
 two-dimensional QCD.  A particular attention is paid to the issues of
 infrared cut-off dependence  and gauge invariance.
 The Renyi  entropy for a single interval, is  given by the rainbow
 dressed quark propagator to order ${\cal O}(N_c)$. The  contributions to order
 ${\cal O}(1)$, are shown to follow from the off-diagonal and off mass-shell mesonic T-matrix,
 with no contribution to the central charge.  The construction is then extended to mesonic
states on the light front, and shown to probe the moments of the partonic PDFs for large LF separations.
 In the vacuum and for small and large intervals,
the spatial entanglement entropy following from the Renyi entropy, is shown to be in agreement with the Ryu-Takayanagi geometrical
 entropy,  using a soft-wall AdS$_3$  model of  two-dimensional QCD.
\end{abstract}

\maketitle

\section{Introduction}
Quantum entanglement is paramount in quantum mechanics. It follows from the fact that quantum states
are mostly superposition states, and two acausally related measurements can be correlated. A quantitative
measure of this correlation is given by the entanglement entropy, with a number of applications in quantum
many body systems and also quantum field theory~\cite{Srednicki:1993im,Calabrese:2004eu,Casini:2005rm,Hastings:2007iok,Calabrese:2009qy}.

The increase interest in entanglement,
especially in lower dimensional systems, is partly motivated by recent developments in quantum information
theory. Of particular interest, is the concept of  entanglement entropy as a measure of quantum information
flow~\cite{le1967proceedings,Bekenstein:1981zz}.
There is a large effort currently underway for a better theoretical and experimental understanding
of entanglement in the nuclear many body problem~\cite{kaufman2016quantum}, the prompt thermalization at RHIC~\cite{Stoffers:2012mn,Qian:2015boa,Berges:2018cny,Florio:2021xvj,Liu:2022ohy},
hadron tomography through DIS~\cite{Kharzeev:2017qzs,Liu:2022hto,Liu:2022ohy}, and parton-parton scattering at low-x~\cite{Stoffers:2012mn,Qian:2014rda,Kharzeev:2017qzs,Shuryak:2017phz,Liu:2018gae,Armesto:2019mna,Dvali:2021ooc}.

Recently, we have shown how  entanglement in longitudinal parton-x, and also in rapidity space or ${\rm ln}\frac 1x$,
can be used to gain more insights on the partonic PDFs (large-x) and structure functions (small-x),  using two-dimensional QCD.
Recall tha  2D QCD is solvable  in the large number of colors limit~\cite{tHooft:1974pnl,Bars:1976nk}. This allows for a
quantitative understanding of the role played by the entanglement entropy, for single meson states, or their stringy form
by resummation along a Regge trajectory. Remarkably, the entanglement entropy carried by a 2D nucleus on the light
front, shows a growth rate with rapidity at the current bound  on quantum information flow.

Spatial entanglement in interacting theories, and especially gauge theories is challenging. The geometrical construction
proposed by Ryu-Takayanagi~\cite{Ryu:2006bv} in the context of a holographic dual gauge theories at large $N_c$ and strong gauge coupling,
in this sense is rather remarkable. In interacting gauge theories with fermions, the dual descriptions are only approximate,
and using them to analyze the entanglement geometrically is interesting, especially if large $N_c$ arguments can be used
for a comparison.

Entanglement in two-dimensional QCD is intricate,  as it involves interacting fermions with a dynamical gauge field.  To address it,
we use the replica construction in {\it real time},  by duplicating  Minkowski space-time $n$ times, and then gluing the duplicates
together, using pertinent
twists of the replicated fermion fields. This procedure makes the ensuing Renyi entropy and its limiting entanglement
entropy, gauge dependent in any dimension. This notwithstanding, both entropies can be evaluated by gauge fixing
both in the continuum or on the lattice. For two-dimensional QCD, we will show that in the regular cut-off gauge,
the large $N_c$ results are found to be in agreement with a soft-wall holographic construction, for very small or very large intervals. For completeness,
we note that  a replica analysis of two-dimensional QCD was  suggested in~\cite{Goykhman:2015sga}, using different
arguments.

The paper is organized as follows: In section~\ref{SECII}, we briefly review the  replica construction of the Renyi entropy,
and its relation to the  entanglement entropy. We will also recall the form of the monodromy matrix that allows for the
gluing of the fermionic replicas. In particular, we will derive a new equal-time representation of the replica partition
function. In section~\ref{SECIII}, we discuss the subtleties related to the gauge symmetry following from the gluing of the
fermions, and why gauge fixing is required across the gluing cut. We will analyse the replica partition function, both
in perturbation theory and in the large $N_c$ limit of 2D QCD, in the light front gauge. In section~\ref{SECIV}, we extend our
replica construction to the spatial entanglement in partonic as well as hadronic states on the light front. For the latter,
the entanglement is controlled by the  moments of the partonic PDFs in 2D QCD. We suggest that these moments can
be extracted from the Renyi entropy for space-like intervals in a fast moving hadron in 4D QCD, using current lattice QCD simulations.
 In section~\ref{SECV},
The leading results of the entanglement
entropy both for small and large intervals, are shown to be compatible with the Ryu-Takayanagi entropy, using a soft-wall
gravity dual to 2D QCD. Our conclusions are in section~\ref{SECVI}.

%


\section{Replica partition function and Renyi entropy}~\label{SECII}
\label{Replica}

Let $\rho$ be the density of a pure state defined in a Hilbert space composed of two complementary regions
$I$ and its complementary $\bar I$. For simplicity, we first focus on spatial regions. The projected or reduced density matrix in
$\bar I$ obtained by tracing over $I$, is~\cite{Calabrese:2009qy,Casini:2005rm}

\bea
\label{1}
\rho_{I}={\rm Tr}_{\bar{I}}\rho
\eea
Although $\rho$ carries zero von Neumann entropy, $\rho_{I}$ does not,

\bea
\label{2}
S=-{\rm Tr}_{I}(\rho_I\,{\rm log}\,\rho_I)
\eea
which is a measure of the quantum entanglement between $I$ and $\bar I$ in $\rho$. To evaluate (\ref{2})
one uses the Replica trick through the Renyi entropy $S_n$

\begin{align}
\label{REN1}
S_n=\frac{1}{1-n}\ln {\rm tr} \rho_{\rm I}^{n}\equiv\frac{1}{1-n} \ln Z_n \ .
\end{align}
If $Z_n$ is analytic in $n$ in a neighborhood of $n=1$ with the Taylor-expanded form:
\begin{align}
\ln Z_n =(n-1)Z^{(1)}+(n-1)^2Z^{(2)}+...  \ ,
\end{align}
then the Shannon entropy or the entanglement entropy can simply be identified as
\begin{align}\label{RENT}
S={\lim_{n\to 1}}S_n=-Z^{(1)}=-{\lim_{n\to 1}}\frac{\partial}{\partial n}\ln Z_n \ .
\end{align}
We now show how to derive the replica partition function using the equal time formulation,
valid for any interacting fermionic theory, in any dimension.

\subsection{Fermionic monodromy}

Using the transfer matrix, Calabrese and Cardy~\cite{Calabrese:2004eu,Calabrese:2009qy} have shown that $Z_n$ for integer value of $n$,  can be rewritten
as an Euclidean path integral with fields living in a replica space. More specifically, a path integral  with $n$ identical
copies of the original Euclidean space, glued together along the single spatial cut corresponding to the region $I$,
 with twisted fermionic boundary conditions.

For a fermionic theory one has for $i=1, ...n$  replicated fermions $\psi_{i}$, each living in its own manifold, this patching corresponds to twisting the fermions in going
from one patch to the  other~\cite{Calabrese:2009qy,Casini:2005rm}

\bea
\label{MONO}
\bigg[{\cal T}_n\bigg]\begin{pmatrix}
\psi_{1}\\
\psi_{2}\\
\vdots\\
\psi_{n}
\end{pmatrix}=
\begin{pmatrix}
0& 1 & 0& \cdots & 0 \\
0& 0& 1&  \cdots & 0\\
\vdots  & \vdots  & \ddots & \vdots & 1  \\
(-1)^{n+1}& 0 & 0& \cdots & 0
\end{pmatrix}
\begin{pmatrix}
\psi_{1}\\
\psi_{2}\\
\vdots\\
\psi_{n}
\end{pmatrix}
\eea
The eigenvalues of the monodromy ${\cal T}_n$ are the n-roots of unity $e^{i2\pi k/n}$ with $k=-\frac{n-1}2, ..., +\frac{n-1}2$.
This amounts to n-multi-valued fermions in a single-cut space $I=[a_1,a_2]$, with each species $\psi_k$ picking a phase $e^{i2\pi k/n}$ phase
in circling the left-edge ($a_1$) of the cut clock-wise, and $e^{-i2\pi k/n}$ in circling the right-edge ($a_2$) of the cut counter-clock-wise.

\subsection{Equal-time representation of $Z_n$}

In a Hamiltonian formulation of the replica in Minkowski signature, the gluing conditions are the new and key  elements to add to the original field theory.
We first consider the case of only fermionic theories with a single spatial cut,  and  the  gluing conditions for the fermions given
in  (\ref{MONO}). To construct the replica partition function for the vaccum of interacting fermions, we  start from the generic off-diagonal matrix element
of the vacuum density matrix $|\Omega\rangle\langle \Omega|$

\begin{align}\label{eq:matrixrho}
\langle \psi_{0^-}|\Omega\rangle \langle \Omega|-\psi_{0^+}\rangle=\langle \Omega|\psi_{0^+}\rangle\langle \psi_{0^-}|\Omega\rangle \ ,
\end{align}
where $|\psi_{0^{\pm }}\rangle$ refer to two generic  fermionic coherent states (their precise relation to the single  space-time cut and  labeling, will be detailed below).
Here  $|\Omega\rangle$ refers to the lowest energy state, prepared using the long time evolution, with the full fermionic Hamiltonian $H(\psi^\dagger, \psi)$
\begin{align}
|\Omega\rangle=e^{-iH[\psi^{\dagger},\psi]\frac{T}{2}(1-i0)}|\psi_{-\infty}\rangle \ .
\end{align}
starting from an arbitrary asymptotic coherent state $|\psi_{-\infty}\rangle$, whose explicit
form is not needed. The additional minus sign in (\ref{eq:matrixrho}) is due to the Grassmannian nature of the states, when moving $\langle \psi_{0^-}|\Omega\rangle$ from left to right.  Also, it is important that the  density matrix $|\Omega \rangle \langle \Omega|$ is bosonic, namely,
when expanded as polynomials in the Grassmannians, the order of each term must be even.

With this in mind, and to proceed to a path integral, we use the decomposition
$$e^{-iHT/2}=e^{-iH\epsilon}e^{-iH\epsilon}e^{-iH\epsilon}...e^{-iH\epsilon}$$
 and insert the completeness relation between any of the two evolution operators
\begin{align}
\label{ONE}
{\bf 1}=\int d\bar\psi_{t} d\psi_t e^{-\bar \psi_t\psi_t}|\psi_t\rangle \langle \psi_t| \ ,
\end{align}
As a result, the matrix element in (\ref{eq:matrixrho}) can be cast in  a standard path-integral  form
\begin{align}
\label{DENSE}
\langle \Omega|\psi_{0^+}\rangle\langle \psi_{0^-}|\Omega\rangle=\int\prod_{t} d\bar \psi_t d\psi_t e^{\sum -  \bar\psi_{t}(\psi_{t}-\psi_{t-1})-i\epsilon H[\bar \psi_{t},\psi_{t-1}]} \ ,
\end{align}
with no $-\bar\psi_{0^+}\psi_{0^-} $ term in the exponent. (\ref{DENSE}) is a path-integral representation of the density matrix in real time, for a single fermion species. To represent the trace,  we need  the completeness relation and the trace formula
\begin{align}
\label{TRACEA}
{\rm Tr} A= \int d\bar \psi d\psi e^{-\bar \psi \psi}\langle -\psi |A|\psi\rangle \ ,
\end{align}
in terms of which, we have
\begin{align}
&{\rm Tr \rho_I^n}=\int \prod_{k=0}^{n-2}d\bar\psi_{k,0^-}d\psi_{k,0^-}e^{-\sum_{k=0}^{n-1}\sum_{x}\bar \psi_{k,0^-}(x)\psi_{k,0^-}(x)}\nonumber \\
&\prod_{k=0}^{n-1}\langle \Omega|-\psi_{k-1,0^-}(x\in I),\psi_{k,0^-}(x\notin I)\rangle\langle \psi_{k,0^-}(x\in I),\psi_{k,0^-}(x\notin I)|\Omega \rangle \ ,
\end{align}
where in the last equation one has made explicit the dependence on $x$ and $\psi_{0-1,0^-}=-\psi_{n-1,0^-}$. The above can then be represented as a path-integral in the replica space time with $n$ replica fermions species and with the gluing boundary condition across the boundary $I$ as indicated explicitly as in the equation above.

\begin{figure}[!h]
\includegraphics[height=7cm]{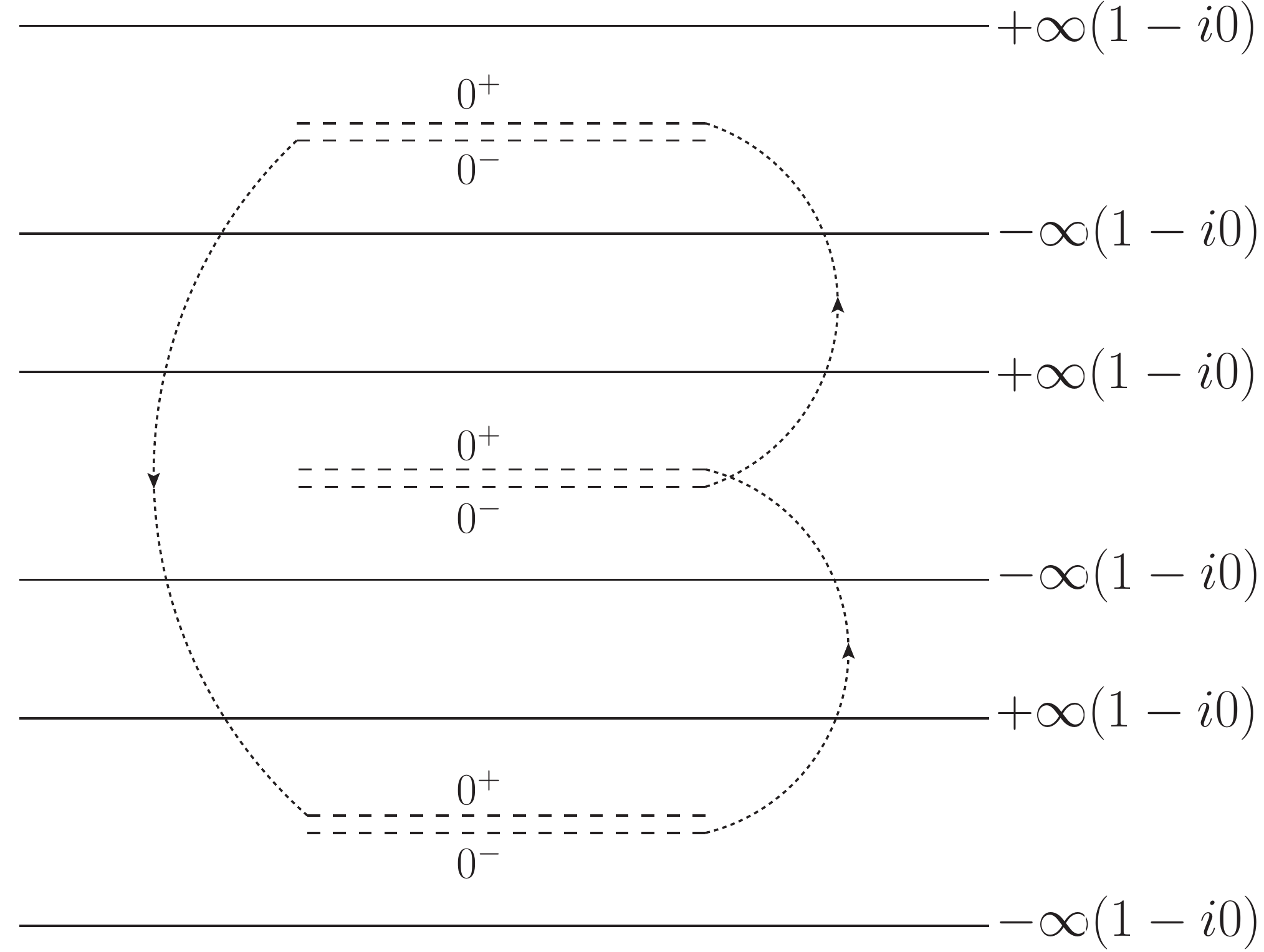}
 \caption{Replica Minkowski space-time for $n=3$. The boundaries for the time evolution at $t=\pm \infty(1-i0)$ are denoted by horizontal solid lines, and the cuts at $t=0^{\pm}$ are denoted by the double dashed lines, in the middle of each replica strip. The fields at the cut for different replica copies,  are glued following  the dotted lines. For a thermal theory with inverse temperature $\beta$, the imaginary time version of the Euclidean space time,  follows a similar construction with $\pm \infty(1-i0)\rightarrow \pm \frac{\beta}{2}$, and periodic (or anti-periodic) boundary conditions at the solid boundaries for each replica strip. }
  \label{fig:replica}
\end{figure}

More specifically, the $n$'th trace can be written as a path integral with $i=0,1,..n-1$ copies of the fermion fields $\psi_{i,t}(x)$. Here $i$ refers to the replica index, $t$ to the time slice and $x$ to the spatial coordination of the Grassmannian.
The twisting across the cut amounts to $\psi_{i,0^+}(x\in I)=-\psi_{i-1,0^-}(x\in I)$ for $i=1,2,...n-1$ and $\psi_{-1,0^+}(x\in I)=-\psi_{n-1,0^-}(x\in I)$,
as illustrated in Fig.~\ref{fig:replica}. Outside the cut, we have $\psi_{i,0^+}(x\notin I)=\psi_{i,0^-}(x\notin I)$. Now, using the charge conservation of the Hamiltonian, one can flip all the Grassmannians for old $i=2k-1$
\begin{align}
&\langle \Omega|-\psi_{2k-2,0^-}(x\in I),\psi_{2k-1}(x\notin I)\rangle \langle \psi_{2k-1,0^-}(x\in I),\psi_{2k-1,0^-}(x\notin I)|\Omega\rangle \nonumber \\
\equiv  &\langle \Omega|\psi_{2k-2,0^-}(x\in I),-\psi_{2k-1}(x\notin I)\rangle \langle -\psi_{2k-1,0^-}(x\in I),-\psi_{2k-1,0^-}(x\notin I)|\Omega\rangle \ ,
\end{align}
and redefine for old $i$ (including $n-1$ if $n$ is even)
\begin{align}
\psi_{2k-1,0^-}(x) \rightarrow -\psi_{2k-1,0^-}(x)  \ .
\end{align}
Clearly, after these transformations, one has the alternative boundary condition $\psi_{i+1,0^+}(x\in I)=\psi_{i,0^-}(x\in I)$ for $i=0,1,2,...n-2$ and $\psi_{1,0^+}(x\in I)=(-1)^{n+1}\psi_{n-1,0^-}(x\in I)$, and for $x \notin I$ one needs no sign change. In terms of the independent variables $\psi_{i,0^-}(x)$, one has in the exponential for fermions along the cut or $x\in I$
\begin{align}
\sum_{i}\bar \psi_{i,1}(x \in I)\psi_{i-1,0^-}(x\in I)-\sum_i \bar \psi_{i,0^-}(x\in I)\bigg(\psi_{i,0^-}(x\in I)-\psi_{i,-1}(x\in I)\bigg)
\end{align}
where $\psi_{-1,0^-}(x\in I)=(-1)^{n+1}\psi_{n-1,0^-}(x\in I)$ according to the boundary condition, in addition to the Hamiltonian term
\begin{align}
-i\epsilon\sum_{i,x} H\bigg[\bar\psi_{i,1}(x),\psi_{i-1,0^-}(x\in I),\psi_{i,0^-}(x\notin I)\bigg]-i\epsilon \sum_i H\bigg[\bar\psi_{i,0^-}(x),\psi_{i,-1}(x)\bigg] \ .
\end{align}
This finishes  the derivation of the replica partition function in real time, with the twisted  boundary conditions across the
 cut $I$, as
illustrated in Fig.~\ref{fig:replica} for $n=3$. Each strip  in Minkowski space-time is cut at the initial times $t=0^\pm$,
which is shown in dashed lines,
with the fermionic field assignments $\psi_{i,0^\pm}(x\in I)$.

To proceed further,  we switch to  the fermionic fields labeled by $k$, that diagonalize the monodromy (\ref{MONO}) for the
original replica fields labeled by $i$
\bea
\label{PSIK}
\psi_{k,t}(x)&=&\frac{1}{\sqrt{n}}\sum^{n-1}_{i=0}e^{-i\frac{2\pi k}{n}i}\psi_{i,t}(x) \ , \\
\psi_{k,t}^{\dagger}(x)&=&\frac{1}{\sqrt{n}}\sum^{n-1}_{i=0}e^{i\frac{2\pi k}{n}i}\psi^{\dagger}_{i,t}(x) \ ,
\eea
at every space-time point, in terms of which the partition function reads
\begin{align}\label{eq:replicafinal}
&\int \prod_{k,x}d\bar\psi_{k,0^-}(x)d\psi_{k,0^-}(x)e^{-\sum_{k,x} \bar \psi_{k,0^-}(x)\psi_{k,0^-}(x)}\nonumber \\
&\langle \psi_{\infty}|e^{-iH[\bar \psi_{k, 0^-},\psi_{k, 0^-}]T/2}|\psi_{k,0^-}(x\in I) e^{\frac{2\pi i k}{n}},\psi_{k,0^-}(x\notin I)\rangle
\langle \psi_{k,0^-}(x)|e^{-iH[\bar \psi_{k, 0^-},\psi_{k, 0^-}]T/2}|\psi_{-\infty}\rangle \ ,
\end{align}
Here
\begin{align}
H[\bar \psi_{k, 0^-},\psi_{k, 0^-}]=\sum_{i}H[\bar \psi_{i,0^-},\psi_{i, 0^-}] \ ,
\end{align}
refers to the Hamiltonian for  $n$ identical copies of the original Hamiltonian,  written in the new variables $\psi_k$, which is seen to satisfy the identity
\begin{align}
|\psi_{k,0^-}(x,\in I) e^{\frac{2\pi i k}{n}},\psi_{k,0^-}(x\notin I)\rangle=e^{i\frac{2\pi k}{n}\sum_{k}\sum_{x\in I}\psi^{\dagger}_{{k, 0^-}}(x)\psi_{{k, 0^-}}(x)}|\psi_{k,0^-}(x)\rangle \ ,
\end{align}
(\ref{eq:replicafinal}) reduces to the expectation value
\begin{align}
\label{KEY1}
\langle \Omega_n| \exp \bigg[i\sum_{k}\frac{2\pi k}{n}\int_{x \in I} dx\psi^{\dagger}_{k, 0^-}(x)\psi_{k, 0^-}(x)\bigg] |\Omega_n \rangle \ ,
\end{align}
$I$ refers to  the cut, and  $|\Omega_n\rangle$ is simply a tensor product of $n$ identical vacua of the original theory, one for each replica copy labeled by $i$.
Note that the exponential, is  the equal-time charge density  in k-space,  conjugate to the replica i-space
\begin{align}
\int_{x \in I} dx\psi^{\dagger}_{k, 0^-}(x)\psi_{k, 0^-}(x)\equiv \int_{x\in I} dx j_{0,k}(x) \ .
\end{align}
From here on, the argument $x$ is short for the equal-time argument $(0^-, x)$ unless specified otherwise.
In terms of the original replica fields labeled by $i$,  (\ref{KEY1}) reads
\begin{align}\label{eq:replicacorre}
Z_n=\langle \Omega_n| \exp \bigg[i\sum_{i,j}\sum_{k}\frac{2\pi k}{n^2}e^{i\frac{2\pi k}{n}(i-j)}\int_{x \in I} dx\psi^{\dagger}_i(x)\psi_j(x)\bigg] |\Omega_n \rangle \ .
\end{align}
 (\ref{eq:replicacorre}) is  the replica partition function or the $n$-trace of the reduced density matrix. It is  an expectation value of equal-time operators in a replica  theory with $n$ copies.

 For a free fermion theory,  (\ref{KEY1})
 reduces to  the result established in~\cite{Casini:2005rm}, based on the interpretation of the replica boundary conditions
  as background magnetic fields  with fluxes $\frac{2\pi k}{n}$. Indeed, analytically continuing (\ref{KEY1}) to Euclidean
  signature,  and using the 2D bosonization relation  $\psi_k^{\dagger}\gamma^{\mu}\psi_k=\frac{1}{\sqrt{\pi}}\epsilon^{\mu\nu}\partial_{\nu} \phi_k$,
  we have
 \bea
&&i\sum_{k}\frac{2\pi k}{n}\int_{x \in I} dx\psi^{\dagger}_{k, 0^-}(x)\psi_{k, 0^-}(x)\equiv \nonumber\\
&&i\sum_k \frac{\sqrt{4\pi}k}{n}\bigg[\phi_k(a_2)-\phi_k(a_1)\bigg]\equiv -i\sum_{k}\int d^2x A^{k}_{\mu}(x)\bar \psi_k(x) \gamma^{\mu}\psi_k(x) \ ,
 \eea
with the replica magnetic fields $$\epsilon_{\mu\nu}\partial^{\mu}A^{k,\nu}(x)=\frac{2\pi k}{n}[\delta^2(x-a_1)-\delta^2(x-a_2)]\,,$$ in agreement with~\cite{Casini:2005rm}. However, our
result~(\ref{eq:replicacorre}) is more general, as it applies to generic interacting fermionic systems in Minkowski signature, including 4-Fermi or gauge interactions.

In sum, we derived an equal time representation for the replica partition function $Z_n=e^{(n-1)S_n}$, for any free or interacting
two dimensional fermionic theory,  along an equal-time space-like cut. It readily  generalizes to any dimensions $D+1$, for any $D$-dimensional space-like region $I$. For free fermions, the above can also be derived using bosonization~~\cite{Armoni:2000uw}, but here we have shown that the same applies to any fermionic theory, with or without interactions. (\ref{KEY1}-\ref{eq:replicacorre}) are the main  results of this section.


\section{Two-dimensional QCD}~\label{SECIII}

Now we proceed to show how the preceding result can be exploited in two-dimensional QCD,  paying particular attention to issues of gauge invariance.
We  present a perturbative analysis of the entanglement entropy for small spatial cuts, followed by a large $N_c$ analysis whatever the size of the cut.

\subsection{Gauge symmetry}

Each of the replicated $n$ copies of two-dimensional QCD,
has local gauge invariance in the corresponding space time, and requires gauge fixing
across each of the replicated cut. More specifically, additional gauge links connecting $i$ to $i+1$ copies in space-time, need to be specified. Indeed,
the exponent in (\ref{eq:replicacorre})
$$\int_{x \in I} dx\psi^{\dagger}_i(x)\psi_j(x)$$
while local in x-space,  is off-diagonal in replica i-space. While gluing the replicated space-times, the gauge transformation from one edge
in the i-patch say at time $0^-$, has to be adjusted so to match the gauge transformation from the other edge in the i+1-patch  at time $0^+$.
This means fixing the gauge along the cut.   In two dimensions we may choose a gauge,  e.g. the axial gauge or temporal gauge, where the only physical degrees of freedom are  fermions,  and then apply the above construction solely to the fermions. The two approaches are not necessarily equivalent. The former in terms of the gauge fields, is  explicitly gauge dependent, while the latter in terms of solely the fermionic fields,  is implicitly gauge dependent through the inverted gauge propagator.  The
elimination procedure of the gauge fields, does not work in higher dimensions.  Finally,  because of local gauge symmetry,  replica partition functions lack in general,
an interpretation as the trace over a reduced density  matrix in a Hilbert space, viewed as a tensor product.

This notwithstanding, we may use (\ref{eq:replicacorre}) in either Minkowski or Euclidean signature as a definition
of $Z_n$, and proceed to evaluate it  either perturbatively,  or non-perturbatively using the planar approximation (alternatively a lattice evaluation).
In all cases, gauge fixing is required. Below, we show that while $Z_n$ and the ensuing Renyi entropy $S_n$, are in general gauge
dependent, the leading contributions at small and large cuts, are gauge independent. The same results, will be shown
to follow from a gauge invariant holographic construction.


\subsection{Perturbative analysis on the light front}

The representation of the fermion replica partition function as an equal-time correlation function, allows generalization to any cut along the direction $n^{\mu}$ in a manifestly invariant manner
\begin{align}
Z_n(x^{\mu}=Ln^{\mu})=\langle \Omega_n|{\cal T}\exp \bigg[\sum_k i\frac{2\pi k}{n}\int_{0}^L ds n^{\mu}\epsilon_{\mu\nu}j^{\nu}_k(sn^{\mu})\bigg]|\Omega_n\rangle \ ,
\end{align}
where $\epsilon_{\mu\nu}j^{\nu}_k(x)$ is the vector current operator for the fermion $\psi_k$. This representation is manifestly Lorentz invariant.
Therefore, the partition function $Z_n(x)$ depends only on the Lorentz invariant length $\sqrt{-x^2}$ of the separation, but not the direction. Furthermore,
assuming that $j^{\nu}_k(x)$ satisfies the standard local commutation relations, one can show that the $Z_n(x)$ should  have the same analyticity properties, in particular the domain of analyticity,  and the $i\epsilon$ prescription as a two point function of local scalar fields.

To proceed, we use the  LC gauge,  and represent the LF time evolution as a path integral, for which we need to evaluate
\begin{align}
\bigg \langle \exp \bigg[\sum_{ij}\sum_k i\frac{2\pi k}{n}e^{i\frac{2\pi k}{n}(i-j)}\int_{a_1^-}^{a_2^-} dx^- \psi_i^{\dagger}(0, x^-)\psi_j(0 ,x^-) \bigg]\bigg \rangle_{\rm int} \ .
\end{align}
But since the equal LF time field is equivalent to a set of free-field, the above is the same as the non-interacting theory. All the vacuum diagrams vanish due to the fact that

\begin{align}
H_{\rm int}|0\rangle_{\rm free}=0 \ .
\end{align}
\begin{figure}[!h]
\includegraphics[height=3cm]{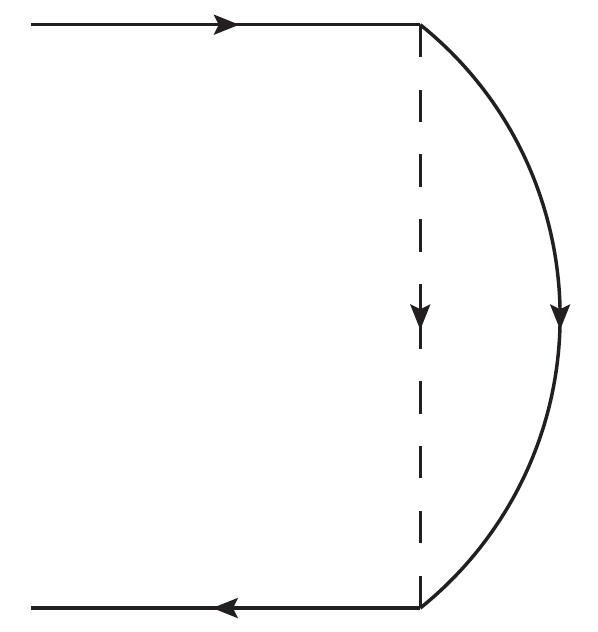}
 \caption{A typical vacuum insertion that vanishes in the LF perturbation theory. }
  \label{fig:vanished}
\end{figure}

\begin{figure}[!h]
\includegraphics[height=3cm]{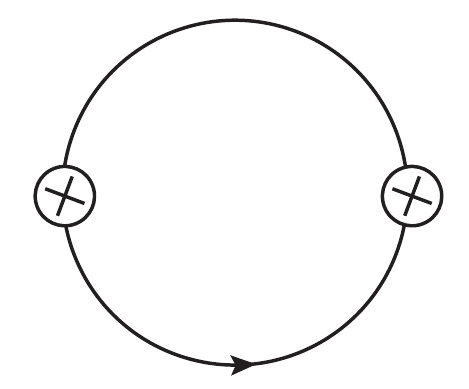}
 \caption{The vacuum polarization contribution to $\ln Z_n$ in Eq.~(\ref{eq:replicafinal}). The crossed dots denote insertions of the operator $\int dx \psi^{\dagger}\psi$. For massless free fermions,  it is the only non-vanishing diagram,  and contributes t the known  $c=\frac{N_c}{3}$. For a  super-renormalizable theory, this is the only diagram that contains a UV divergence. }
  \label{fig:bubble}
\end{figure}

The representation as a correlation function,  allows a perturbative expansion using standard Feynman rules. For a free fermion, this reproduces the well known result. Indeed, if one consider $\ln Z_n$, then only the connected diagrams will contribute
\begin{align}
\ln Z_n(L)=\sum \text{connected diagrams with insertions of $\int dx \psi^{\dagger}\psi$} \ .
\end{align}
For a free fermion, this means loops with arbitrary numbers of  $\bar \psi_k \gamma^0 \psi_k$ insertions. However, due to the absence of anomalies for any fermion loop with more than three fermion propagators, an application of the vector and axial Ward identities, shows that all loops (with more than three insertions) vanish. The only non-vanishing diagram is the vacuum polarization diagram shown in Fig.~\ref{fig:bubble}, at the origin of the 2D axial anomaly. A direct calculation  leads to the standard central charge $\frac{N_c}{3}$.

More specifically, the vacuum polarization diagram in~Fig.~\ref{fig:bubble} contributes as
\begin{align}
\ln Z_n|_{\rm bubble}=\frac{N_c}{2}\sum_{k}\bigg(\frac{2\pi k}{n}\bigg)^2 \int \frac{d^2p}{(2\pi)^2}\Pi^{00}(p)\frac{2-2\cos p^z L}{(p^z)^2} \ ,
\end{align}
for a massive fermion one has the well known vacuum polarization in $2D$
\begin{align}
\Pi^{00}(p)=\frac{(p^z)^2}{\pi}\int_{0}^1 dx \frac{1}{p^2+\frac{m^2}{x(1-x)}} \ ,
\end{align}
with the result
\begin{align}
\ln Z_n|_{\rm bubble}=-\frac{(n^2-1)N_c}{12n}\int_{0}^1 dx \int^{\infty}_{-\infty} dp_z \frac{1-\cos p_zL}{\sqrt{p_z^2+\frac{m^2}{x(1-x)}}} \ .
\end{align}
The first term diverges in  the UV. Using the  UV regulator $$1-\cos p_zL \rightarrow \cos p_za-\cos p_zL \,,$$  the result is
\begin{align}
\ln Z_n|_{\rm bubble}=-\frac{(n^2-1)N_c}{6n}\int_{0}^1 dx \bigg[K_0\bigg(\frac{ma}{\sqrt{x(1-x)}}\bigg)-K_0\bigg(\frac{mL}{\sqrt{x(1-x)}}\bigg)\bigg] \ ,
\end{align}
with the Renyi entropy (\ref{REN1}) in the form
\bea
\label{REN2}
S_n=\frac{(n+1)N_c}{6n} \int_{0}^1 dx \bigg[K_0\bigg(\frac{ma}{\sqrt{x(1-x)}}\bigg)-K_0\bigg(\frac{mL}{\sqrt{x(1-x)}}\bigg)\bigg] \rightarrow \frac {N_c}3 {\rm ln}\bigg(\frac{L}a\bigg)
\eea
The rightmost result follows in the massless limit ($m\rightarrow 0$), for $n=1$.  The L-dependent central charge is
\begin{align}
c_n(L)=L\frac{dS_n}{dL}=\frac{(n+1)N_c}{6n}\int_{0}^1 dx\frac{mL K_1(\frac{mL}{\sqrt{x(1-x)}})}{\sqrt{x(1-x)}} \ ,
\end{align}
 which is seen to decay exponentially as $N_ce^{-2mL}$, at large $L$. The Renyi entropy (\ref{REN2}) at large $L$, is dominated by the
constant UV contribution
\bea
\label{REN3}
\frac{(n+1)N_c}{6n}\bigg( \int_{0}^1 dx K_0\bigg(\frac{ma}{\sqrt{x(1-x)}}\bigg)+{\cal O}(e^{-2mL})\bigg)\rightarrow
\frac{(n+1)N_c}{6n}\bigg( {\rm ln}\bigg(\frac{1}{ma}\bigg) +{\cal O}(e^{-2mL})\bigg)\nonumber\\
\eea

Since  the interaction is super-renormalizable (valid also for 2D QED), any diagram with interactions vertices will be less singular than the vacuum polarization diagram. In other words, they are  UV free, and contribute ${\cal O}\big(g^{2n}L^{2n}\big)$ at short distances. The dominant contribution at small $L$, is therefore
\begin{align}\label{eq:entanglefree}
S(L)=\frac{N_c}{3}\ln \frac{L}{a}+{\cal O}\big(g^{2}L^{2}\big) \ .
\end{align}
On the other hand, we  expect exponential decay with $L$ at large $L$, for massive  fermions.

Finally, we note that for  two disjoint intervals,  the above formalism allows the calculation of the so-called mutual information,
\begin{align}
&\ln Z_n(L_1\cup L_2)-\ln Z_n(L_1)-\ln Z_n(L_2)\nonumber \\
&=\sum \text{connected diagrams with both insertions in $L_1$ and $L_2$ } \ .
\end{align}
Since the distance between $L_1$ and $L_2$ are non-zero, the diagrams have a natural UV cutoff and will be convergent. This applies even to super-renormalizable theories (Gross-Neveu) after coupling constant renormalization.

\subsection{Summing planar contributions with replicas: counting $n-1$ }

In the large $N_c$ limit, the leading contribution is again dominated by a single planar fermion-loop with possible insertions of the charge-operators. We are only interested in the leading $n-1$ contributions that lead to the entanglement entropy.  Here we present a power-counting argument that eliminates most of the diagrams.

Notice that the insertions of the $\int dx \psi^{\dagger}\psi$ operators in each of the fermion propagator, have the generic structure
\begin{align}
{\cal G}_{i,j}(p,p')=\delta_{ij}G_{0}(p,p')+\sum_{m=1}^{\infty} G_m(p,p') A^m_{ij} \ ,
\end{align}
where $p$ and $p'$ denotes the incoming and outgoing momenta, and
\begin{align}
A^n_{ij}=\sum_{k=-\frac{n-1}{2}}^{\frac{n-1}{2}}e^{-i\frac{2\pi k}{n}(i-j)}\bigg(\frac{k}{n}\bigg)^n \ ,
\end{align}
is an  ij-matrix in replica space, with eigenvalues $(\frac{k}{n})^n$. For any diagram, the $n$ dependence
follows from the trace over matrices formed by $A$, depending on the locations and numbers of the insertions.

\begin{figure}[!h]
\includegraphics[height=6cm]{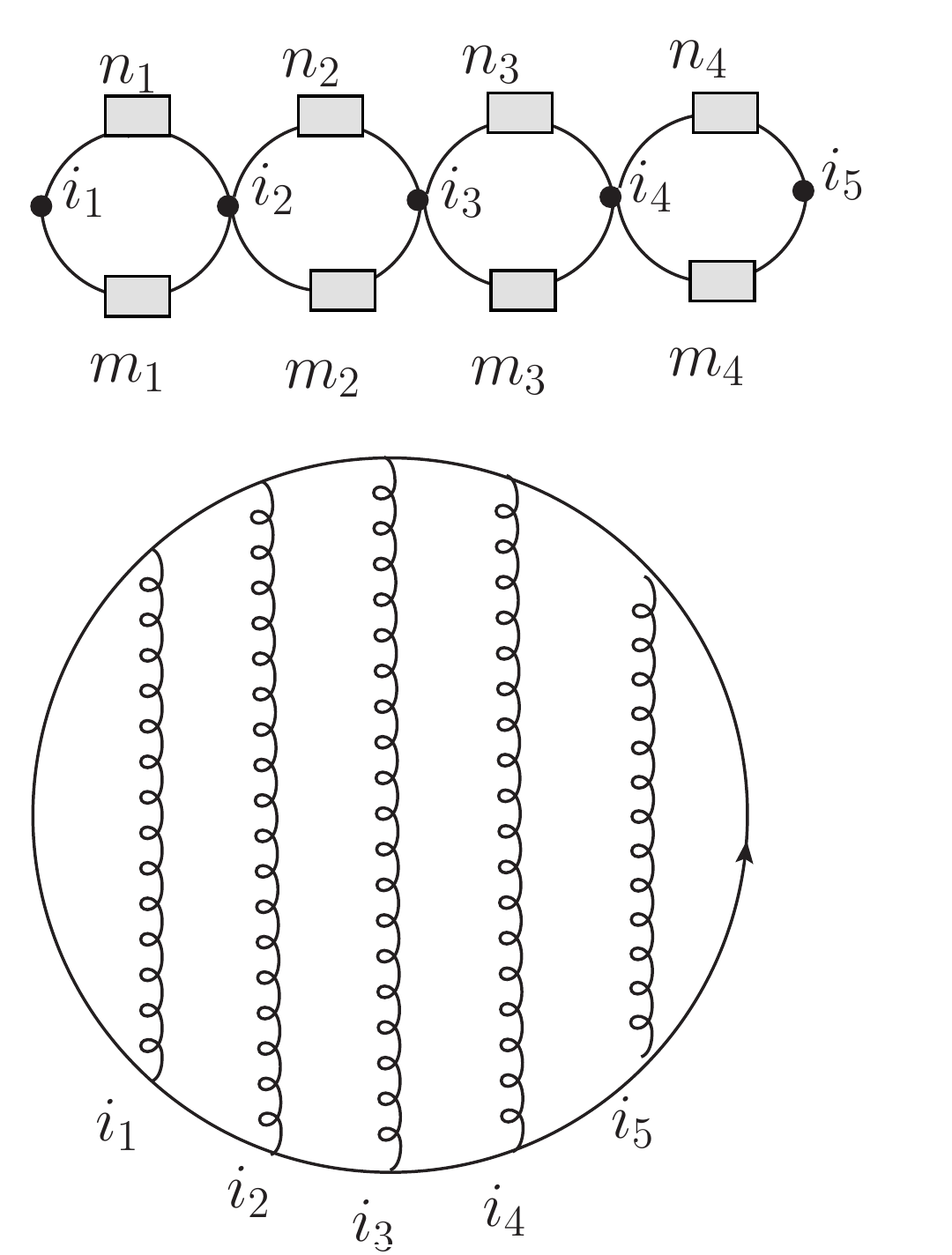}
 \caption{A generic replica-color structure of a planar diagram  that contributes to ${\cal O}(N_c)$.
 The upper diagram follows by inserting the replica fluxes $(n_i, m_i)$ in the lower diagram, in each of the lines
 between the gluonic exchanges. The dotted 4-Fermi interactions labeled by the replica index $i$ in the upper diagram,
 is short for the integrated  gluon exchange from the lower diagram in 2D.}
  \label{fig:colorsingle}
\end{figure}

Now consider the generic replica-color structure shown in Figure.~\ref{fig:colorsingle}. Inside a single fermion loop there is a ladder formed by $N$ instantaneous gluons.
Let's now make insertions on the fermion propagators. The number of powers of $A$ on each rung is labelled by $(n_i, m_i)$ where $i=0,1,2..N$. Lets show that there exits only a single $i$ in which one of the $(n_i,m_i)$ can be non-vanishing. Indeed, one can go from the left side, by summing over $i_1$ one obtain
\begin{align}
A^{n_1+m_1}_{ii}\propto  \sum_{k=-\frac{n-1}{2}}^{\frac{n-1}{2}}\frac{k^{n_1+m_1}}{n^{n_1+m_1}} \ ,
\end{align}
which is independent of $i$, and is always proportional to $(n-1)$ as long as $n_1+m_1\ne 0$. Therefore, if $n_1+m_1 \ne 0$, no other insertions are allowed. Otherwise one obtain $\delta_{i_1i_2}$ and go to $(n_2,m_2)$. Continue this way the assertion is confirmed.

Given the rules above, it is not hard to find  the diagrams that are leading in $n-1$. Indeed, a generic planar diagram can be obtained from  Fig.~\ref{fig:colorsingle} by inserting rainbow-like 1PI diagrams,  on each of the fermion propagator. If the operator insertions are outside such rainbows, then the replica-color structure remains the same, and the above argument applies.  Specifically, for the $i$-ring with the  insertion numbers $(n_i,m_i)$ possibly non-zero, one may  add rainbows between the insertions, without changing the counting in $n-1$.  Moreover, if the insertions are inside such rainbows, then by moving the legs of the gluons along the contour, one can view the gluons inside the rainbow, as forming a ladder. The other gluons that used to be a ladder, become rainbows. In this way we are again reduced to the previous case.

\subsection{Order ${\cal O}(N_c)$ contribution}

\begin{figure}[!h]
\includegraphics[height=6cm]{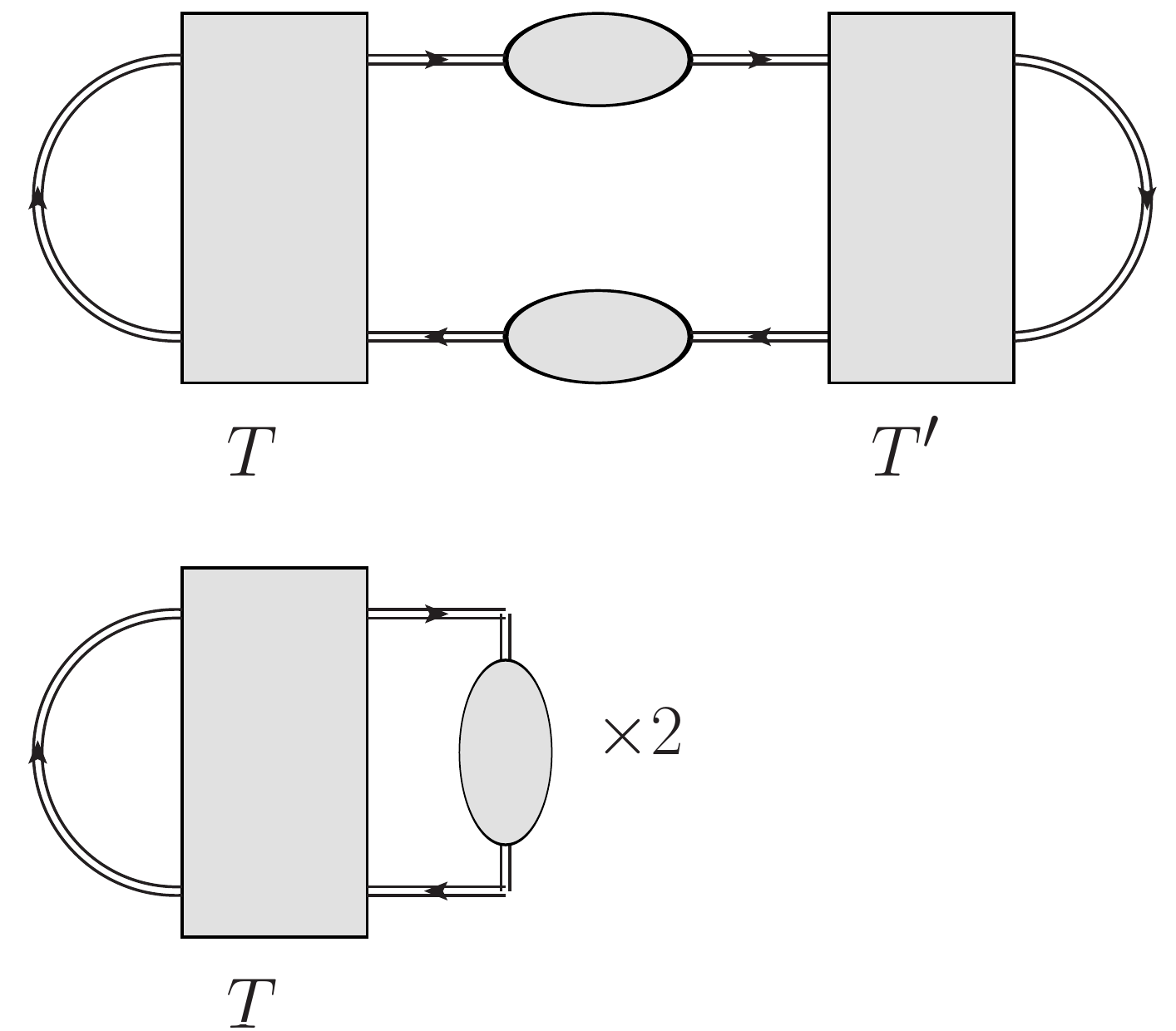}
 \caption{The single-loop contribution. The shaded boxes are the planar two to two amplitudes and the insertions are located at the shaded circles.  Notice that between two insertions there can be arbitrary numbers of rainbows.}
  \label{fig:singleloop}
\end{figure}

The diagrams that are leading have the topological structure shown in~Fig~\ref{fig:singleloop}.
In the upper diagram above, at least one of $T$ and $T'$ is non-trivial. If one of $T$ and $T'$ is trivial, then the first diagram  reduces to the lower one. However, notice that in these cases the $T$ and $T'$ themselves can be viewed as forming rainbows, therefore the above diagrams are really equivalent to the following: arbitrary number of operators inserted in a fermion-loop with arbitrary number equal or grater than $1$ of rainbows inserted along the fermion propagators between them. When combined with the diagram without any rainbow insertions, the fermion propagator between the operator insertions resummes to the dressed one.

With this in mind,  the leading $N_c$ contribution to the entanglement entropy is actually equivalent to that of a free fermion,
but  with a rainbow dressed propagator
\begin{align}
\label{SR}
S_{N_c}=N_cS(\langle \psi(x-y) \bar \psi(0)\rangle_{\rm Rainbow}) \ ,
\end{align}
Here $S(G_{\rm Rainbow}(x-y))$ denotes the entanglement entropy for a free fermion,  with a rainbow dressed propagator~\cite{Einhorn:1976uz,Frishman:1975fd}
\bea
\label{GR}
G_{\rm Rainbow}((x-y))=\int \frac{d^2p}{(2\pi)^2}e^{-ip(x-y)}
\frac{p^+\gamma^++(p^-+\frac {g^2N_c}{2\pi p^+}-Ag^2N_c{\rm sign}(p^+) )\gamma^-+m}{p^2-m^2+\frac 1\pi g^2N_c-A  g^2N_c |p^+|}\nonumber\\
\eea
with $A$ a gauge parameter.

The fact that (\ref{SR}) through (\ref{GR}) depends on $A$,
means that in general, the entanglement entropy in a gauge theory,
 is inherently gauge dependent, even after  the elimination of the gauge degrees of freedom
 in 2D QCD as we discussed earlier.  We note that
 $'$t Hooft originally identified $A=\frac 1{\epsilon^-}$ with an infrared cutoff~\cite{tHooft:1974pnl}, for which its removal from (\ref{GR}) will cause the  contribution (\ref{SR}) to vanish.  However, this is a particular gauge choice. In the  $A=0$ gauge (regular cutoff prescription)~\cite{Einhorn:1976uz}, the rainbow resummation in (\ref{GR}) is non-vanishing,  with
 a renormalized squared mass $\tilde m^2=m^2-g^2N_c/\pi\geq 0$.

Since the gauge dependent part of the self-energy does not change the short distance behavior, the small $L$ behavior of the resummed entanglement entropy,
in the planar approximation, is still dominated by the vacuum polarization diagram. It is gauge invariant (independent of $A$),
 and is equal to  $\frac{N_c}{3}\ln \frac{L}{a}$.
 This result is reminiscent of the current-current 2-point function which is given by the free fermion loop and of order $N_c$~\cite{Callan:1975ps}, an illustration of  parton-hadron duality in 2D QCD.  For $\tilde m^2>0$, the asymptotics of the central charge is seen to vanish as $N_ce^{-2\tilde m L}$, with the Renyi entropy dominated by the constant  UV contribution (\ref{REN3}) at large $L$,  which is also gauge independent! These results are   unaffected by the ${\cal O}(1)$ contributions as we  discuss below.

Finally, we note that the case $m=0$ is pathological with $\tilde m^2<0$ tachyonic. In this case, the left and right hand fermions decouple, with the
fermionic propagator for the right-hand particle unchanged, while for the left particle it changes to
\begin{align}
\label{GTACH}
G^+(z)=e^{-ig^2N_cA|z|}\gamma^-{\rm sign}(z)\int_{0}^{\infty} \frac{dk^+}{4\pi}e^{-ik^+|z|-i\frac{g^2N_c-i0}{\pi k^+}|z|} \ .
\end{align}
At long distance, (\ref{GTACH})  decays only polynomially as ${1}/{z^{\frac{3}{2}}}$,
 and the ensuing entanglement will decay also polynomially. On the other hand, since the right-hand fermion remains free,
 it will contribute only $\frac{N_c}{6} \ln \frac La$ at long distances!

\begin{figure}[!h]
\includegraphics[height=6cm]{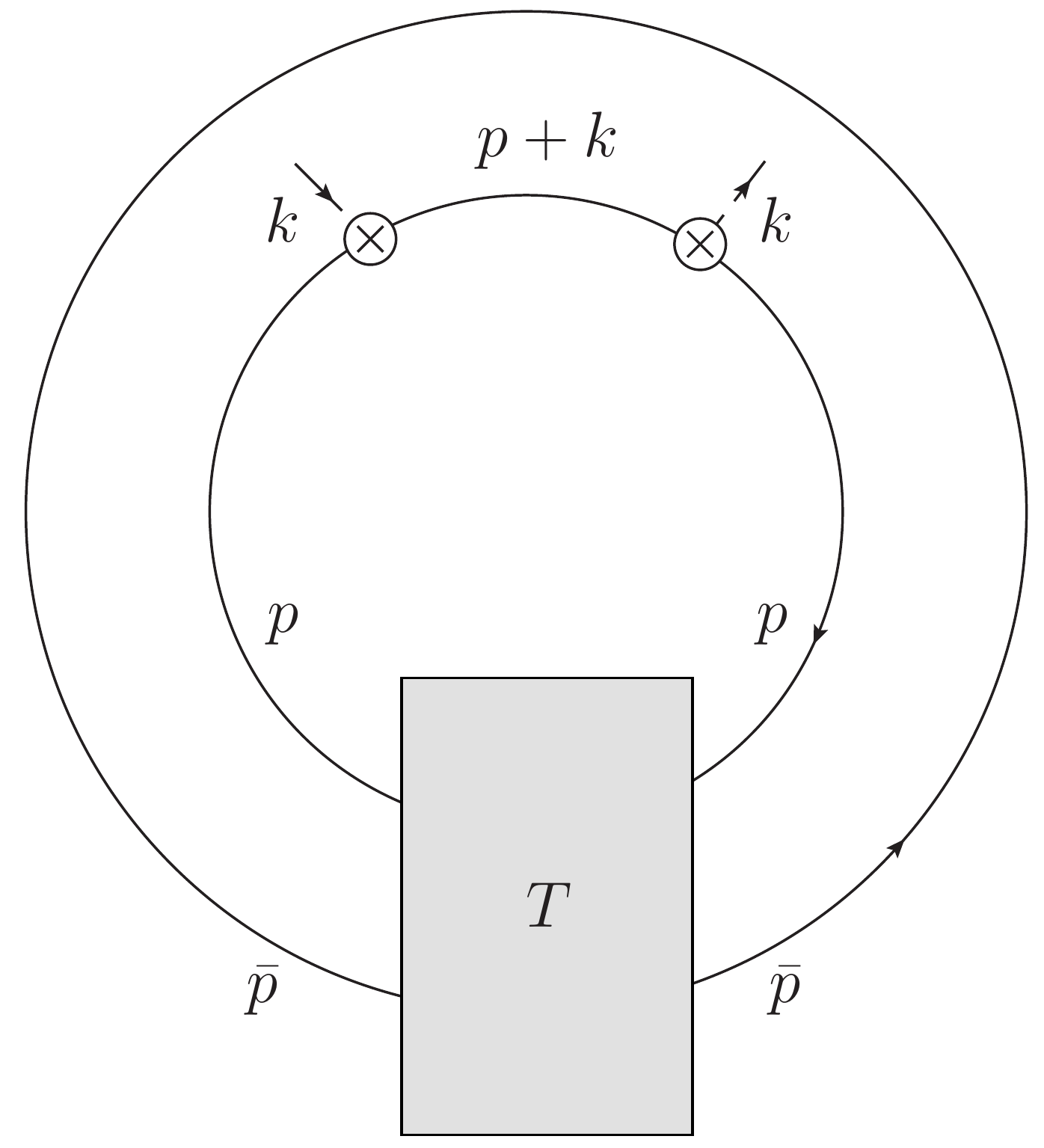}
 \caption{The first non-vanishing double-loop contribution to $Z_n$. The shaded box is the amputated two-body planar amplitude. The crossed circles are  the insertions of $\psi^{\dagger}\psi$.}
  \label{fig:double}
\end{figure}

\subsection{Order ${\cal O}(1)$ contribution}

The ${\cal O}(1)$ contributions in the planar approximation, resums the independent mesonic contributions to the entanglement entropy.
The meson spectrum contains a would-be Goldstone mode, that may shift the large distance part of the central charge
from  $\frac {N_c}3$ to $\frac{N_c}3+\frac 13$. We now show that this is not the case.

\begin{figure}[!h]
\includegraphics[height=4cm]{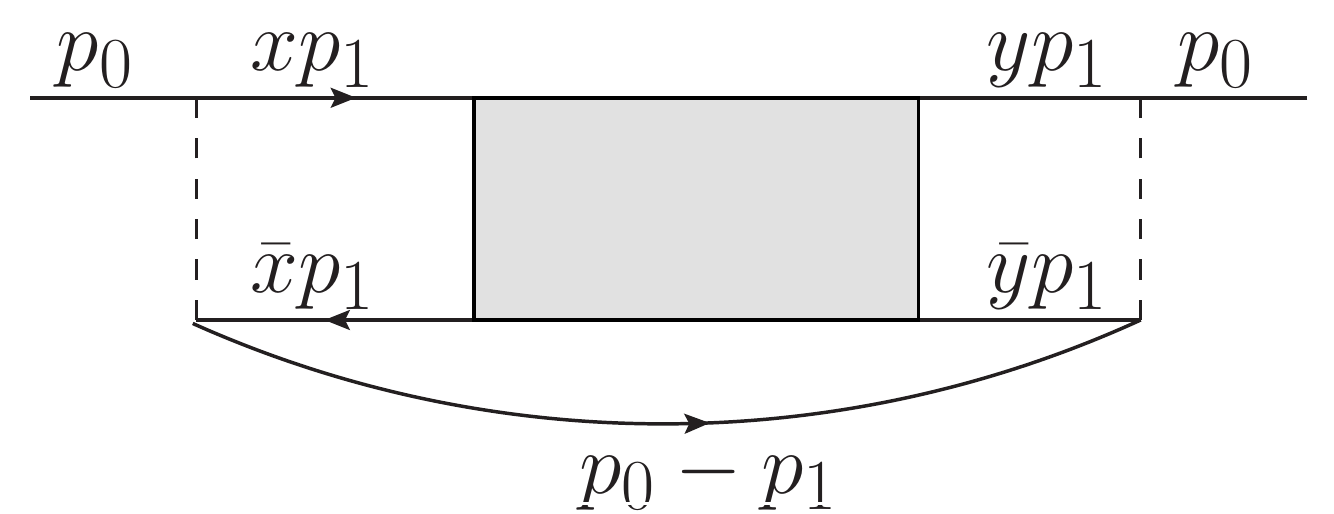}
 \caption{The LF  diagram for $\tilde T(p_0^+,p_0^-)$ where $p_0^+$ is positive (it is negative for the flipped antiquark line). The shaded box represents the equal incoming-outgoing LF time $T$ matrix, and the dashed line represents the  instantaneous gluon at equal LF time.}
  \label{fig:doubleLF}
\end{figure}

The ${\cal O}(1)$ contribution is illustrated in Fig.~\ref{fig:double}. In momentum space, it translates to
\begin{align}
\ln Z_n|_{\rm double}=2\times \frac{1}{2}\sum_{k}\bigg(\frac{2\pi k}{n}\bigg)^2 \int \frac{d^2k}{(2\pi)^2}\tilde \Pi^{00}(k)\frac{2-2\cos k^z L}{(k^z)^2} \ ,
\end{align}
with $\tilde \Pi^{00}(k)$ given by
\begin{align}
\tilde \Pi^{00}(k)=\int \frac{d^2p}{(2\pi)^2} {\rm tr} S(p)\gamma^0 S(p+k)\gamma^0 S(p)\tilde T(p) \ ,
\end{align}
and
\begin{align}
\tilde T^{aa'}_{\alpha\alpha'}(p)=\int \frac{d^2\bar p}{(2\pi)}T^{aa';bb}_{\alpha\alpha';\beta\beta'}(p,\bar p)S_{\beta'\beta}(\bar p) \ .
\end{align}
Note that only the {\it forward but off-mass shell} part of the $T$ matrix is needed. In light cone gauge, $\tilde T$ follows from~Fig.~\ref{fig:doubleLF}.

To evaluate this, one first notices that the equal incoming-outgoing time $T$ matrix in LF gauge is simply given by
\begin{align}
&T(r^+,r^-,x,y)=\frac{g^2}{(r^+)^2}\bigg(\frac{\pi r^2}{g^2N_c}-\frac{\gamma-1}{x}-\frac{\gamma-1}{\bar x}\bigg)\delta(x-y) \nonumber \\
&-\frac{g^2}{(r^+)^2}\bigg(\frac{\pi r^2}{g^2N_c}-\frac{\gamma-1}{x}-\frac{\gamma-1}{\bar x}\bigg)\bigg(\frac{\pi r^2}{g^2N_c}-\frac{\gamma-1}{y}-\frac{\gamma-1}{\bar y}\bigg)G(x,y,r^2) \ ,
\end{align}
with $\gamma=\pi m^2/g^2N_c$. The incoming $+$ component momenta for the quark and the antiquark,  are $xr^+$ and $\bar xr^+$, and the total incoming LF energy are $r^-$. The mesonic Green function $G(x,y,r^2)$ can be written in terms of the $^\prime$t Hooft  LF wavefunctions $\phi_n(x)$ for mesons with squared masses
$m_n^2/g^2N_c \sim n\pi$ (large $n$)
\begin{align}
G(x,y,r^2)=\sum_{n}\frac{\varphi_n(x)\varphi_n(y)}{\frac{\pi r^2}{g^2N_c}-\frac{\pi m_n^2}{g^2N_c}} \ .
\end{align}
Thus, $\tilde T$ can be calculated as
\begin{align}
\tilde T(p_0)=\frac{(g^2N_c)^2}{\pi N_c}  \sum_n \int_{0}^1 dx \int_{0}^1 dy\int dp_1^- &\int_{0}^{p_0^+} \frac{dp_1^+}{(2\pi)^2}\frac{\varphi_n(x)\varphi_n(y)}{(p_0^+-xp_1^+)^2(p_0^+-yp_1^+)^2}\nonumber \\
&\times \frac{(p_0-p_1)^+}{(p_0-p_1)^2-m^2+\frac{g^2N_c}{\pi}}\frac{(p_1^+)^2}{(p_1)^2-m_n^2} \ .
\end{align}
in the gauge with $A=0$ (regular cut-off prescription).
The above integral is convergent at $p_0^+=p_1^+$ only if $\varphi_n(x)\sim x^{\beta}$ near the edges with $0<\beta<1$. The above can be calculated as
\begin{align}
&p_0^+\tilde T(p_0) \equiv \tilde \Sigma(p_0^2)\nonumber \\
&=\frac{(g^2N_c)^2}{\pi^2 N_c }\sum_{n}\int_{0}^1 dxdydz \frac{\varphi_n(x)\varphi_n(y)}{(1-xz)^2(1-yz)^2}\frac{z}{p_0^2-\frac{m_n^2}{z}-\frac{m^2-\frac{g^2N_c}{\pi}}{1-z}+i0} \ .
\end{align}
This is actually the order ${1}/{N_c}$ correction to the quark-self energy. For an estimation, when $m^2=0$, there exists zero mass solution to the 't Hooft equation with $m_n=0$, in this case the contribution reads
\begin{align}
\Sigma(p_0^2)\sim \frac{(g^2N_c)^2}{\pi^2 N_c }\int_{0}^1 dxdydz \frac{\varphi_0(x)\varphi_0(y)}{(1-xz)^2(1-yz)^2}\frac{z}{p_0^2+\frac{\frac{g^2N_c}{\pi}}{1-z}+i0}  \ .
\end{align}
If one use $\phi_0=1$, the integral diverges logarithmically  near $z=1$. For  small but finite $m$, the contribution
is of  order ${\sqrt{g^2 N_c}}/{m}$. When resummed into the fermion propagator, we have
\begin{align}
S(p_0)=\frac{p_0^+}{p_0^2-m^2+\frac{g^2N_c}{\pi}+\Sigma(p_0^2)} \ .
\end{align}
in the $A=0$ gauge (regular cutoff prescription). A rerun of the preceding arguments yields a central charge $\frac {N_c}3$, with no additional
$\frac{1}{3}$ contribution from the would-be Golstone mode at long distances.

\section{Spatial entanglement in excited states}~\label{SECIV}

The present analysis can be  generalized to any excited state $|N\rangle$. Using the
pertinent interpolating fields to create the excited meson or baryon states,  (\ref{eq:replicacorre}) readily generalizes to
\begin{align}
\label{ZNL}
Z_{N_n}(L)=\langle N_n|{\cal T}\exp \bigg[\sum_k i\frac{2\pi k}{n}\int_{0}^L ds n^{\mu}\epsilon_{\mu\nu}j^{\nu}_k(sn^{\mu})\bigg]|N_n\rangle  \ ,
\end{align}
where $|N_n\rangle$ is  a  tensor product of $|N\rangle$, one for each replica copy,
\begin{align}
|N_n\rangle=\bigotimes_{i=0}^{n-1}|N\rangle_i \ .
\end{align}
Moreover, if we  choose $n^{\mu}$ to be along the LF$^-$ direction, then (\ref{ZNL}) is reminiscent of LF  parton distribution functions.

\subsection{Free parton on the light front}

For a free fermion state of longitudinal momentum $P^+$ or $|N\rangle=b_{P^+}^{\dagger}|\Omega\rangle$,  the contributions for different $k$ factorize,
\begin{align}
\label{FF1}
\ln Z_n=(1-n)S_n+\sum_{k=-\frac{n-1}{2}}^{\frac{n-1}{2}}\ln \bigg[\int^{\Lambda^-/2}_{-\Lambda^-/2} \frac{dxdy}{2\pi\Lambda^-}\frac{ie^{-i(x-y)}}{x-y+i0}\bigg(\frac{(x-\lambda+i0)(y-i0)}{(y-\lambda-i0)(x+i0)}\bigg)^{\frac{k}{n}}\bigg] \ .
\end{align}
Here  $R^-$ is the box size along LF$^-$, and $\Lambda^-=P^+R^-$ and $\lambda=P^+L^-$ the invariant lengths. In deriving (\ref{FF1}), we used the bosonized
representation for the fermion field $\psi_k\sim e^{i\phi_k}$ in (\ref{ZNL}).  In the large LF box limit with $L^-/R^-\ll 1$, the kernel in (\ref{FF1}) can be reduced,
\begin{align}
\label{ZFF2}
&\ln Z_n-(1-n)S_n=-\frac{4\lambda}{\Lambda^-}\sum_{k=-\frac{n-1}{2}}^{\frac{n-1}{2}}\sin^2\frac{k\pi}{n}\int_{0}^{1} \frac{dxdy}{2\pi}\bigg[\frac{(1-x)y}{(1-y)x}\bigg]^{\frac{k}{n}}\frac{\sin \lambda(x-y)}{x-y}\ .
\end{align}
The details are in Appendix~\ref{DETAILS}. The entanglement entropy follows by  performing the $n\rightarrow 1$ limit in (\ref{ZFF2}), using the formula~\cite{Casini:2005rm}
\begin{align}
\lim_{n\rightarrow 1}\frac{1}{1-n}\sum_{k=-\frac{n-1}{2}}^{\frac{n-1}{2}}\sin^2\frac{k\pi}{n}z^{\frac{k}{n}}\sim-\lim_{n\rightarrow 1}\frac{2\pi^2(n-1)}{4\pi^2(n-1)^2+(z-1)^2}=-\pi^2\delta(z-1)\ ,
\end{align}
with the result
\begin{align}
\label{ZFF3}
S=S(L^-)+\frac{4\pi^2\lambda}{\Lambda^-}\int_{0}^{1} \frac{dxdy}{2\pi}\delta(x-y)y(1-y)\frac{\sin \lambda(x-y)}{x-y}=S(L^-)+\frac{\pi\lambda^2}{3\Lambda^-} \ .
\end{align}
 $S(L^-)$ is the vacuum entanglement entropy discussed earlier.
For large LF$^-$ intervals with invariant length $\lambda=P^+L^-$, the entanglement entropy of a free fermion on the LF is of order $\frac{\lambda^2}{\Lambda^-}$.
For small intervals, it is dominated by the Logarithmic contribution from the vacuum in $S(L^-)$. In particular, for a free  fermionic parton with the least  longitudinal
momentum $P^+=\frac{2\pi}{R^-}$,  (\ref{ZFF3})  simplifies to
\bea
S=S(L^-)+\frac{2\pi^2}3 \bigg(\frac{L^-}{R^-}\bigg)^{2}\, .
\eea
The additional contribution
is the entanglement entropy for a primary state in a free conformal field theory~\cite{Berganza:2011mh} with $h=1$ and $\bar h=0$.

\subsection{Free meson on the light front}

Consider  a bound meson state on the LF,  with  longitudinal momentum $P^+$,
\begin{align}\label{MESONn}
|l\rangle=B_{l,P^+}^{\dagger}|\Omega\rangle\equiv \frac{1}{(\Lambda^-)^{\frac{1}{2}}}\int d\lambda_1 d\lambda_2 \varphi_l(\lambda_1,\lambda_2)\psi^{\dagger}(\lambda_1)\psi(\lambda_2)|\Omega\rangle \ ,
\end{align}
with the coordinate space light-front wave function (LFWF)
\begin{align}
\varphi_l(\lambda_1,\lambda_2)=\frac{1}{(2\pi)^2}\int_{0}^1 dx e^{-i\lambda_1x-i\lambda_2(1-x)}\varphi_l(x) \ .
\end{align}
and the normalization  $\int dx |\varphi_l(x)|^2=1$. In the replica states constructed from (\ref{MESONn}), the replica partition function is
\bea
\label{ZJN}
Z_n(l)= \langle \Omega|\prod_{j=0}^{n-1}B_{l,j}\exp\bigg[\sum_k i\frac{2\pi k}{n}\int_{0}^{\lambda} d\lambda'\psi_k^{\dagger}(\lambda')\psi_k(\lambda')\bigg]\prod_{j'=0}^{n-1} B_{l,j'}^\dagger |\Omega \rangle \ .
\eea
The correponding entanglement entropy to  leading order in $1/\Lambda^-$, is of the form (\ref{ZFF3}).
More specifically, it is  proportional to $\lambda^2$, but dressed but the second moments of the quark/antiquark PDFs
\begin{align}\label{eq:meson}
S=S(L^-)+\frac{\pi \lambda^2}{3\Lambda^-}\big(\langle x_q^2\rangle_l+\langle x_{\bar q}^2\rangle_l\big) +{\cal O}\bigg(\frac 1{\Lambda^{-2}}\bigg)\ ,
\end{align}
where
\bea
\langle x_{q}^2\rangle_l=\int_0^1 dx\, x^2 \,|\varphi_l(x)|^2 \ , \qquad\qquad \langle x_{\bar q}^2\rangle_l=\int_0^1 dx\, \bar{x}^2\,|\varphi_l(x)|^2 \ ,
\eea
 are the second moments of the quark and antiquark PDFs.
 The higher and even moments of the PDFs are suppressed by further powers of $1/\Lambda^-$, in the entanglement entropy (\ref{eq:meson}).
For the meson state in the Schwinger model, each of the second moment is $\frac{1}{3}$.

\begin{figure}[!h]
\includegraphics[height=8cm]{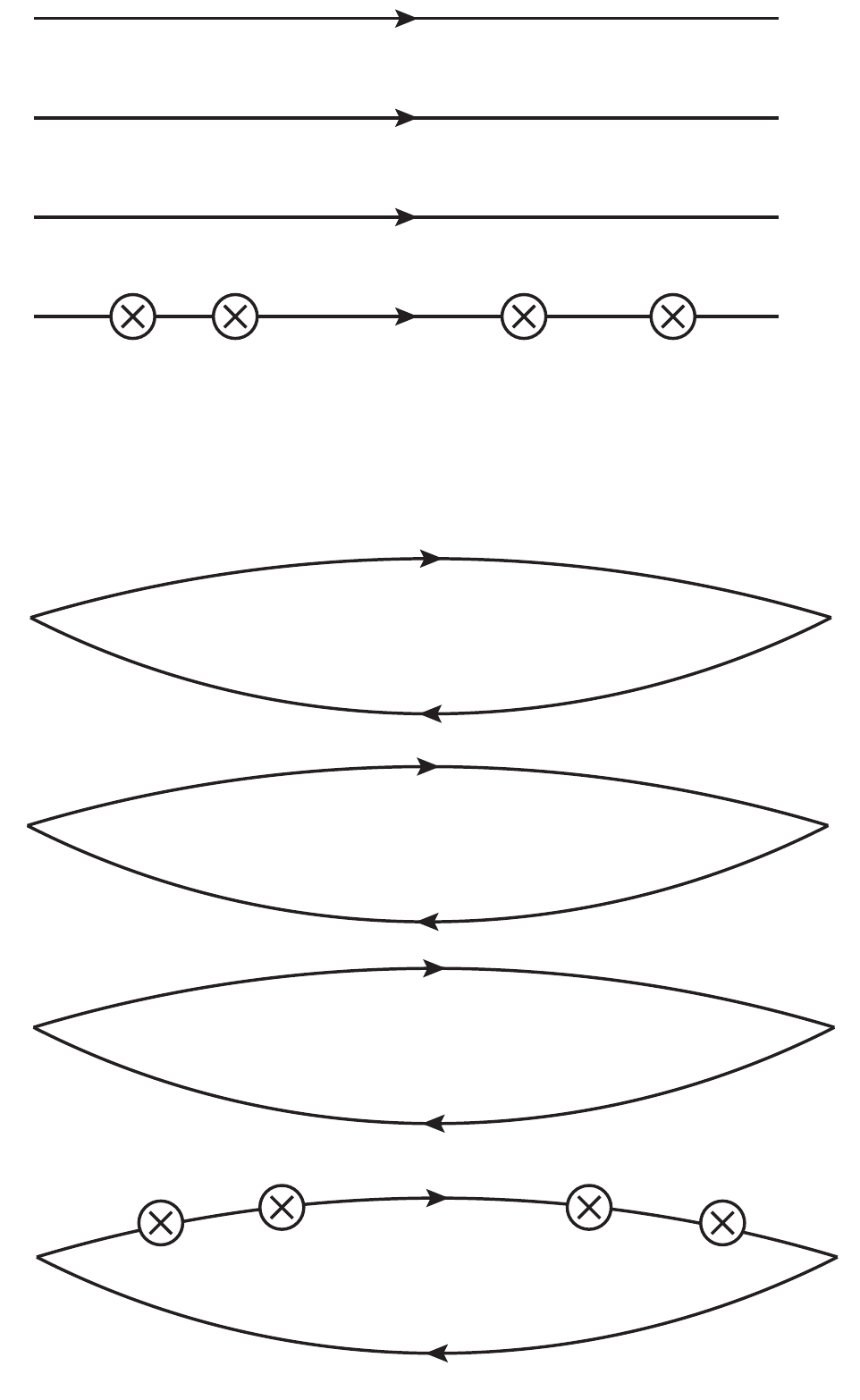}
 \caption{The leading ${1}/{\Lambda^-}$ contribution to the spatial entanglement for  n-replicated fermion  (upper) and n-replicated meson states (lower). The crossed circles denote the vector current operator insertions. To leading order in ${1}/{\Lambda^-}$, $n-1$ pairs of the replicated external states contract with themselves, leaving only one pair for the vector current insertion.
To leading order in $n-1$, no additional insertion is needed.}
  \label{fig:singlee}
\end{figure}

To derive (\ref{eq:meson}), it is best to use a diagrammatic analysis of (\ref{ZJN}) as illustrated in~Fig.~\ref{fig:singlee}.
The disconnected bubbles where the meson operators contract among themselves,  exponentiate and contribute to the vacuum state entanglement.
So we need to  consider only the connected diagrams where the combination $\psi_i^{\dagger}\psi_i$ from the external state,  contracts with $\psi_k^{\dagger}\psi_k$
from the vector operator in the exponent.  We now note that each time a $\psi_i^{\dagger}\psi_i$ from the external state contracts with $\psi_k^{\dagger}\psi_k$ from the operator
insertion,   a suppression factor ${1}/{\Lambda^-}$ arises. Hence, the leading ${1}/{\Lambda^-}$ contribution consists of  $n-1$ pairs of external state contracted among themselves,
with the remaining pair contracted with $\psi_k^{\dagger}\psi_k$ from the vector operator insertion.

For a replicated  fermion in~Fig.~\ref{fig:singlee} (top), this contribution is the trace over the $i$-fields, which readily converts to the sum over  the k-fields.
This  reproduces the second term in \ref{ZFF1}. The extra $-1$ corresponds to the subtraction of the term with no insertions.

This observation extends to the replicated meson state as well. The leading contribution is  shown in Fig.~\ref{fig:singlee} (bottom). For a generic $n$, the operators can be inserted simultaneously on the fermion/antifermion lines. To obtain the linear contribution in $n-1$, one  needs the insertions exclusively on either the fermion, or the antifermion legs, but not both. In this case one reproduces the above free fermion contributions, but weighted over the LFWF of the meson,
\begin{align}
S-S(L^-)=\frac{1}{\Lambda^-}\int_{0}^1 dx |\varphi_n(x)|^2\bigg(F_{\rm single}(x\lambda)+F_{\rm single}(\bar x\lambda)\bigg)\ ,
\end{align}
where $F_{\rm single}(\lambda)=\frac{\lambda^2\pi}{3}$ is the  fermion contribution. This is (\ref{eq:meson}),  and concludes our derivation.  One should mention that although the above derivation is for a free replicated meson state, it can be extended to 2D QCD, using the large $N_c$ power-counting methods detailed above.

We  note that for space-like cuts, the replica partition function (\ref{ZJN}) can be regarded as a meson-meson correlation function,
with replicated fermionic vector charge insertions. In the limit where the meson sources are asymptotically separated, it is in general a
function of the form $Z_n(P\cdot L,  P\cdot R, L^2, R^2)$, and can be probed on an Euclidean lattice in the same spirit as the
quasi-PDF approach in~\cite{Ji:2013dva,Ji:2014gla},
for parton densities. For say large $P^z$  and fixed
spatial cut $L^z<R^z={}^4\sqrt{V_4}$, the second moment of the quark PDF in a meson state can be read from the coefficient of the
Renyi entropy that scales like $1/P^zR^z$.

\subsection{Coherent meson state on the light front}

In a general bosonic coherent state  $$|\xi\rangle =e^{-\frac{|\xi^2|}{2}-\xi B_l^{\dagger}}|\Omega\rangle$$
constructed using (\ref{MESONn}) with  $\xi$ complex valued,  the replica partition function is
\begin{align}
\label{ZNB}
Z_n(\xi)=\prod_k Z_k(\xi) =\prod_k
e^{-|\xi^2|}\langle \Omega|\exp [-\xi^* B_l]\exp\bigg[i\frac{2\pi k}{n}\int_{0}^{\lambda} d\lambda'\psi^{\dagger}(\lambda')\psi(\lambda')\bigg] \exp [-\xi B_l^\dagger]  |\Omega \rangle \ .
\end{align}
For 2D QCD the reduction of (\ref{ZNB}) in terms of the LFWF $\varphi_l(x)$ is straightforward, but tedious.
This construction maybe used to probe for many-body correlations. (\ref{ZNB}) simplifies considerably for 2D QED or the Schwinger model.
Indeed, for the latter $B_l$ is nothing but the bosonized field, and (\ref{ZNB}) can be reduced by bosonization to
\begin{align}
\ln Z_k(\xi)=\ln Z_k-\frac{2\xi k}{n}\frac{\sqrt{2\pi}}{\sqrt{P^+L^-}}\sin\lambda \ .
\end{align}
where $Z_k$ is the vacuum contribution. After summing over $k$, all the $k$ dependent terms cancel out, with only the vacuum contribution remaining. For the Schwinger model, the bosonic coherent state has the same LF-spatial entanglement as that of the vacuum.

\section{Holographic dual construction}~\label{SECV}

In this section, we will construct a soft wall holographic dual to two-dimensional QCD, using the bottom up approach.
 Using the Ryu-Takayanagi proposal~\cite{Ryu:2006bv},
we will derive the entanglement entropy geometrically. We will illustrate the derivation, by recalling the construction for
two-dimensional CFT with an AdS$_3$ gravity dual, and then extend it to the non-conformal case of two-dimensional
QCD using soft-wall AdS$_3$.

\subsection{AdS$_3$}

Two-dimensional conformal theories map onto AdS$_3$, with a central charge $c=3R/2G_3$, with
$R$ the radius of AdS$_3$, and $G_3$ the bulk Newton gravitational constant. In this regime, the entanglement entropy  for the
single spatial cut  $L=|a_1-a_2|$, can be read in bulk using the Ryu-Takayanagi proposal~\cite{Ryu:2006bv}

\bea
\label{SALARGE}
S=\frac{ \gamma_L}{4G_3}\rightarrow \frac {c}3\,{\rm log}\bigg(\frac Ra {\rm sin}\bigg(\frac {\pi L}R\bigg)\bigg)
\eea
with $\gamma_L$ the length of the bulk AdS$_3$ geodesic.
In two dimensions $G_3=g_sl_s$ and $R/G_3=(R/l_s)/g_s$,
with the string length $l_s$. The string coupling  is $g_s\sim 1/N_c$, with the $1/N_c$ universal from
the genus expansion.
For conformal  fermions in the {fundamental} representation, we expect  $R/l_s=\#>1$ (below \# is of order 1), with $c=N_c$.

In Poincare coordinates with line element

\bea
ds^2=\frac{R^2}{z^2}\bigg(-dt^2+dx^2+dz^2\bigg)
\eea
the geodesic  is a semi-circle  ${\dot{x}}^{2}+{\dot{z}}^{ 2}=(L/2)^2$,

\bea
(x(s),z(s))=\frac L2({\rm cos}\,s, {\rm sin}\,s)
\eea
sustained by the single-cut end-points $\pm L/2$,  on  the Minkowski boundary  at $z=a\ll L$ (range $2a/L\leq s\leq \pi-2a/L$).
The  geometric
entanglement entropy is the length of the geodesic in Planck units

\bea
\label{SMALL}
S=\frac {1}{4G_3}\int_{2a/L}^{\pi/2}\,ds\,\sqrt{g_{MN}{\dot{x}}^{M}{\dot{x}}^{N}}=
\frac {R}{2G_3}\int_{2a/L}^{\pi/2}\,\frac{ds}{\rm sin\,s}= \frac {R}{2G_3}{\rm log}\bigg(\frac {\pi L}{a}\bigg)
\eea

\begin{figure}[!htb]
\includegraphics[height=6cm,width=6cm]{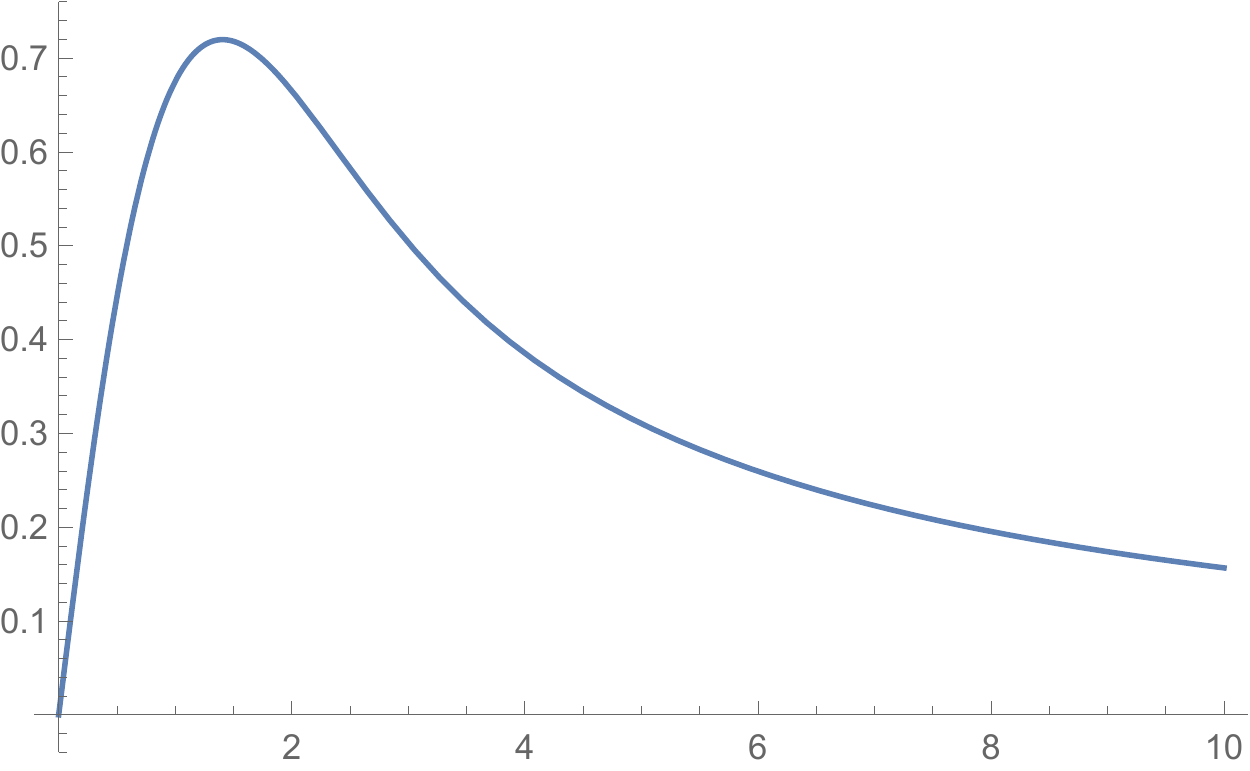}
\caption{$F(\tilde z_m)$ in (\ref{FLA}) as a function of $\tilde z_m$, with a maximum  at $\tilde z_m \sim 1.4$. }
\label{fig:softwall}
\end{figure}

\subsection{Soft-wall AdS$_3$}

Assuming now the geometry is controlled by a soft-wall AdS$_3$
\begin{align}
ds^2=\frac{e^{-\kappa^2z^2}}{z^2}(dz^2+dx^2-dt^2) \ ,
\end{align}
where $\kappa^2 \propto g^2N_c$ plays the role of the ``string tension''. The minimal surface is parameterized by
\begin{align}
(x,z,t)=(x(s),z(s),0) \ ,
\end{align}
where $0\le s\le 1$. The 2-dimensional bulk action   is
\begin{align}
S=\int ds \frac{e^{\frac{-\kappa^2z^2}{2}}}{z}\sqrt{\dot x^2+\dot z^2} \ ,
\end{align}
for which the  minimal surface  can be chosen to satisfy
\begin{align}
\frac{\dot x}{\sqrt{\dot x^2+\dot z^2}}\frac{e^{-\frac{\kappa^2z^2}{2}}}{z}=\alpha \ , \\
\dot x^2+\dot z^2=\beta^2 \ ,
\end{align}
which leads to
\begin{align}
\dot z^2=\alpha^2\beta^2(z_m^2e^{\kappa^2z_m^2}-z^2e^{\kappa^2z^2}) \ ,\\
\dot x=\alpha\beta ze^{\frac{\kappa^2}{2}} \ ,
\end{align}
Here $z_m$ is the maximal value of $z$ attained at $s=\frac{1}{2}$, which satisfies
\begin{align}
\label{LL1}
\frac{L}{2}=\int_{0}^1 ds \dot x =\int_{0}^{z_m} dz\frac{ze^{\frac{\kappa^2 z^2}{2}}}{\sqrt{z_m^2e^{\kappa^2z_m^2}-z^2e^{\kappa^2z^2}}} \ .
\end{align}
For small $L$ (\ref{LL1}) reproduces the circular solution in AdS$_3$ discussed above.
For large $L$, we define  $\tilde z_m=z_m\kappa$ and $\tilde L=L\kappa$ , so that
\begin{align}
\label{FLA}
\frac{\tilde L}{2}=\tilde z_m \int_{0}^1dt \frac{t}{\sqrt{e^{\tilde z_m^2(1-t^2)}-t^2}}\equiv F(\tilde z_m).
\end{align}
For small $\tilde z_m \ll 1$, $F(\tilde z_m)=\tilde z_m$ and the circular solution follows. However,
a maximum develops for $\tilde z_m\sim 1.4$,  so that $F(\tilde z_m)\le 0.72$. The connected solutions only exist for
small $L$, at strong 't Hooft coupling
\begin{align}
L\le \frac{1.44}{\kappa} \sim \frac{1.44}{\sqrt{g^2 N_c}}\ .
\end{align}
For large $L$ the minimal surface cannot be smoothly connected to  the small $L$ solution. A similar observation was also made for
D-branes in higher dimensions, where at large $L$ the solution was argued to be made of  two disjoints in-falling geodesics~\cite{Klebanov:2007ws}. For the soft-wall model,
 this  disconnected geometry  can be approximated by

\bea
\label{LARGE}
S\approx 2\times\frac{R}{4G_3}\int_a^{z_m}\frac {dz}z=\frac{R}{2G_3}{\rm log}\bigg(\frac{z_m}a\bigg)
\eea
The net entanglement entropy is a competition between the circular (\ref{SMALL}) and disjoint (\ref{LARGE}) geometries,

\bea
\label{DSA}
\Delta S=S(\kappa L\ll 1)-S(\kappa L\gg 1)=\frac{R}{2G_3}{\rm log}\bigg(\frac {\pi L}{z_m}\bigg)\rightarrow  \frac{N_c}3\, {\rm log}\bigg(\frac {\pi L}{z_m}\bigg)
\eea
The Ryu-Takayanagi entropies  for small (\ref{SMALL}) and large (\ref{LARGE}) spatial cuts,
are in agreement with the perturbative Renyi  entropy (\ref{REN3}), and its non-perturbative
analogue  $m\rightarrow \tilde m$ at large $N_c$,  respectively.

This interpolation between
a connected surface for small cuts, and a disconnected surface for large cuts is similar to
the observation put forth in~\cite{Klebanov:2007ws}, for several holographic constructions dual to
4D conformal and confining gauge theories. However, the chief difference in our case stems from the fact that 2D QCD at large $N_c$,
confines at all distance scales. The geometrical change we observed,  is not related to a Hagedorn-like growth
in the confined meson spectrum as argued in 4D QCD in~\cite{Klebanov:2007ws}, as there is none in 2D, but is rather a reflection of parton-hadron duality for small
intervals in 2D QCD.

\section{Conclusions}~\label{SECVI}

We have shown how to extend the replica construction to Minkowski space-time signature, and use it to derive
a general formula for the replica partition function in the vacuum state. Our result applies to a large class of
interacting theories with fermions with or without gauge fields, for any space-time cut and in arbitrary dimensions.
When analytically continued to Euclidean signature, our result can be explicitly reduced to the standard result,
using bosonization.

In the presence of gauge interactions, spatial entanglement as described by our replica partition function, is
in general gauge dependent, a result of gluing fermionic fields valued in different replica strips along the spatial
cut.  However, the ensuing Renyi entropy for small or large cuts can still exhibit gauge independent contributions.
We have shown that this is the case in two-dimensional QCD.

For small space-like cuts, the Renyi entropy was shown to follow from the  charge density correlation function,
which is fixed at short distance by the 2D axial anomaly. The central charge is $\frac {N_c}3$ and gauge independent.
At large distances, the perturbative arguments break down.  Using the planar expansion, we showed that the leading
${\cal O}(N_c)$ contribution is tied to the rainbow dressed quark propagator, which is explicitly gauge fixing dependent.
However, for large cuts, this contribution vanishes exponentially with the distance $L$, leaving behind only the gauge
independent UV constant contribution.  The mesonic ${\cal O}(1)$ contributions do not change this result.

Our results are not limited to the vacuum state. We have shown that spatial entanglement on the light front can be
extended to any hadron state, with minimal changes to our central result for the replica partition function. The result
is reminiscent of LF wavefunctions, which shows a direct relationship between the Renyi entropy of an excited hadron,
and its parton distribution on the light front. Conversely and for space-like intervals, the even moments of the quark
PDFs in a hadron state in 2D QCD, can be extracted from the Renyi entropy at large momentum. This observation
extends to 4D QCD both in the continuum, and on an Euclidean lattice.

Using a bottom-up soft-wall model for 2D QCD in AdS$_3$, we have shown that the Ryu-Takayanagi
geometrical entropy, interpolates between the known conformal AdS$_3$ result for a small spatial cut, and a constant
but UV sensitive result for a large spatial cut. This result is  in total agreement with the Renyi entropy,
following from our new replica construction. Although  2D QCD at large $N_c$, is not conformal at all distance scales, the agreement
with the conformal AdS$_3$ result for small intervals, illustrates the parton-hadron duality at work in theories with
confinement.

\vskip 1cm
{\bf Acknowledgements}

This work is supported by the Office of Science, U.S. Department of Energy under Contract No. DE-FG-88ER40388, and by the Priority Research Area SciMat under the program Excellence Initiative - Research University at the Jagiellonian University in Krak\'{o}w.

\appendix

\section{Details in the kernel reduction}~\label{DETAILS}

In  the large $\Lambda^-$ limit one can split the kernel (\ref{FF1}) into
\begin{align}
&\int^{\Lambda^-/2}_{-\Lambda^-/2} \frac{dxdy}{2\pi\Lambda^-}\frac{ie^{-i(x-y)}}{x-y+i0}\frac{(x-\lambda+i0)(y-i0)}{(y-\lambda-i0)(x+i0)}\bigg)^{\frac{k}{n}}\nonumber \\
&=1+\frac{1}{\Lambda^-}\int^{\Lambda^-/2}_{-\Lambda^-/2} \frac{dxdy}{2\pi}\frac{ie^{-i(x-y)}}{x-y+i0}\bigg[\bigg(\frac{(x-\lambda+i0)(y-i0)}{(y-\lambda-i0)(x+i0)}\bigg)^{\frac{k}{n}}-1\bigg] \ ,
\end{align}
and
\begin{align}
\label{ZFF1}
\ln Z_n=(1-n)S_n+\frac{1}{\Lambda^-}\sum_{k=-\frac{n-1}{2}}^{\frac{n-1}{2}}\int^{\Lambda^-/2}_{-\Lambda^-/2} \frac{dxdy}{2\pi}\frac{ie^{-i(x-y)}}{x-y+i0}\bigg[\bigg(\frac{(x-\lambda+i0)(y-i0)}{(y-\lambda-i0)(x+i0)}\bigg)^{\frac{k}{n}}-1\bigg] \ .
\end{align}
The reduction of (\ref{ZFF1}) follows by noting that the bracket  is of the form
\begin{align}
F(z)=\ln (z-\lambda)-\ln z \ ,
\end{align}
with  a branch cut along $[0,\lambda]$ with discontinuity $2\pi i$. The ensuing integral follows by contour
\begin{align}
&\int^{\Lambda^-/2}_{-\Lambda^-/2} \frac{dxdy}{2\pi}\frac{ie^{-i(x-y)}}{x-y+i0}\bigg[\bigg(\frac{(x-\lambda+i0)(y-i0)}{(y-\lambda-i0)(x+i0)}\bigg)^{\frac{k}{n}}-1\bigg]\nonumber \\
&=\int^{\Lambda^-/2}_{-\Lambda^-/2} \frac{dxdy}{2\pi}\frac{ie^{-i(x-y)}}{x-y+i0}e^{\frac{k}{n}F(x+i0)-\frac{k}{n}F(y-i0)} \nonumber \\
&=\int^{\Lambda^-/2}_{-\Lambda^-/2} \frac{dxdy}{2\pi}\frac{ie^{-i(x-y)}}{x-y+i0}\bigg[e^{\frac{k}{n}F(x-i0)-\frac{k}{n}F(y-i0)}+\delta A(x)e^{-\frac{k}{n}F(y-i0)}\bigg]\nonumber \\
&=\Lambda^-+\int^{\Lambda^-/2}_{-\Lambda^-/2} \frac{dxdy}{2\pi}\frac{ie^{-i(x-y)}}{x-y+i0}\delta A(x)e^{-\frac{k}{n}F(y-i0)} \nonumber \\
&=\Lambda^-+\int^{\Lambda^-/2}_{-\Lambda^-/2} \frac{dxdy}{2\pi}\frac{ie^{-i(x-y)}}{x-y+i0}\delta A(x)\delta B(y)+\int^{\Lambda^-/2}_{-\Lambda^-/2} dx \delta A(x)e^{-\frac{k}{n}F(x+i0)} \ ,
\end{align}
with
\begin{align}
\delta A(x)=e^{\frac{k}{n}F(x-i0)}-e^{\frac{k}{n}F(x+i0)}=(e^{-\frac{2\pi k}{n}i}-1)e^{\frac{k}{n}F(x+i0)}\theta(x)\theta(\lambda-x) \ ,\\
\delta B(y)=e^{-\frac{k}{n}F(y+i0)}-e^{-\frac{k}{n}F(y-i0)}=(1-e^{\frac{2\pi k}{n}i})e^{-\frac{k}{n}F(y+i0)}\theta(y)\theta(\lambda-y) \ .
\end{align}
Using the PP-assignment $\frac{1}{x-y+i0}={\rm PV}.\frac{1}{x-y}-i\pi\delta(x-y)$, we obtain (\ref{ZFF2}).

\bibliography{ENT}

\end{document}